\documentstyle[11pt,epsfig]{article}
\def\edth{\;\raise1.0pt\hbox{$'$}\hskip-6pt\partial\;}
\def\baredth{\;\overline{\raise1.0pt\hbox{$'$}\hskip-6pt
\partial}\;}
\def\gsim{~\rlap{$>$}{\lower 1.0ex\hbox{$\sim$}}}

\def\be{\begin{equation}}
\def\ee{\end{equation}}
\def\ba{\begin{eqnarray}}
\def\ea{\end{eqnarray}}

\newcommand{\fr}[2]{\frac{#1}{#2}}

\begin{document}

\title{Exact analytic solution of sub-horizon scales linear perturbation for general dark energy models.}

\author{Seokcheon Lee$^{\,1,2}$}

\maketitle

$^1${\it Institute of Physics, Academia Sinica,
Taipei, Taiwan 11529, R.O.C.}

$^2${\it Leung Center for Cosmology and Particle Astrophysics, National Taiwan University, Taipei, Taiwan 10617, R.O.C.}

\begin{abstract}
Three decades ago Heath found the integral form of the exact analytic growing mode solution of linear density perturbation $\delta$ in sub-horizon scales including the cosmological constant or the curvature term. Interestingly, we are able to obtain the analytic solution for general dark energy models with the constant equation of state $\omega_{de}$. We compare the correct analytic growing mode solution $\delta$ with the text book solution $\delta^{D}$. Indeed, both solutions are equal to each other when $\omega_{de} = -1$. We also able to extend this solution for the specific form of time varying $\omega_{de}$.

\end{abstract}

The background evolution equations in a flat Friedmann-Robertson-Walker universe ($\rho_m + \rho_{de} = \rho_{cr}$) are \ba H^2 \equiv \Bigl(\fr{\dot{a}}{a}\Bigr)^2 &=& \fr{8\pi G}{3}(\rho_{m} + \rho_{de}) = \fr{8 \pi G}{3} \rho_{cr} \, , \label{H} \\ 2 \fr{\ddot{a}}{a} + \Bigl(\fr{\dot{a}}{a}\Bigr)^2 &=& - 8 \pi G \omega_{de} \rho_{de} \, , \label{dotH} \ea where $\omega_{de}$ is the equation of state (eos) of dark energy, $\rho_{m}$ and $\rho_{de}$ are the energy densities of the matter and the dark energy, respectively. We consider the constant $\omega_{de}$.
The sub-horizon scale linear perturbation equation with respect to the scale factor $a$ are given in the reference \cite{Bonnor}, \be \fr{d^2 \delta}{da^2} + \Biggl( \fr{d \ln H}{d a} + \fr{3}{a} \Biggr) \fr{d \delta}{d a} - \fr{4 \pi G \rho_{m}}{(aH)^2} \delta = 0 \, . \label{dadelta} \ee  We rewrite the above equation \be \fr{d^2 \delta}{d x^2} + \Bigl(\fr{1}{2} - \fr{3}{2} \omega_{de} \Omega_{de} \Bigr) \fr{d \delta}{d x} - \fr{3}{2} (1 - \Omega_{de}) \delta = 0 \, , \label{dhx} \ee where $x = \ln a$ and $\Omega_{de} = \Bigl(\fr{\Omega_{m}^{0}}{\Omega_{de}^{0}} a^{3 \omega_{de}} + 1 \Bigr)^{-1} \equiv (Y + 1)^{-1}$. We are able to find the exact analytic growing mode solution of $\delta$ for any value of the constant $\omega_{de}$. After replacing new parameter $Y$ in the equation (\ref{dhx}), we have \be Y \fr{d^2 \delta}{dY^2} + \Bigl[1 + \fr{1}{6 \omega_{de}} - \fr{1}{2(Y+1)} \Bigr] \fr{d \delta}{dY} - \Bigl[\fr{1}{6 \omega_{de}^2 Y} - \fr{1}{6 \omega_{de} Y(Y+1)} \Bigr] \delta = 0 \, . \label{dhY} \ee Now we try $\delta(Y) = c Y^{\alpha} B(Y)$ because it is the most general combination of the solution for the above equation (\ref{dhY}). Now we replace $\delta$ into the above equation (\ref{dhY}) to get, \ba && Y (1 + Y) \fr{d^2B}{dY^2} + \Biggl[ \fr{3}{2} - \fr{1}{6 \omega_{de}} + \Bigl( 2 - \fr{1}{6 \omega_{de}} \Bigr) Y \Biggr] \fr{d B}{dY} + \Bigl( \fr{(3 \omega_{de} +2)(\omega_{de} -1)}{12 \omega_{de}^2} \Bigr) B = 0 \,\, \nonumber \\ && {\rm when} \,\,\, \, \alpha = \fr{1}{2} - \fr{1}{6 \omega_{de}} \,\, .  \label{dskY} \ea There are two alternative ways to make the above equation as the Hypergeometric differential equation, $Y = -X$ or $1+Y = X$. The complete solution of the above equation becomes \ba B(Y) &=& c_{1} F [\fr{1}{2} - \fr{1}{2\omega_{de}}, \fr{1}{2} + \fr{1}{3 \omega_{de}}, \fr{3}{2} - \fr{1}{6 \omega_{de}}, -Y] \nonumber \\ && \, + \, c_{2} Y^{\fr{1-3\omega_{de}}{6 \omega_{de}}} F[-\fr{1}{3\omega_{de}}, \fr{1}{2 \omega_{de}}, \fr{1}{2} + \fr{1}{6 \omega_{de}}, -Y] \, , \label{B} \ea
where $F$ is the hypergeometric function.
Thus, the full analytic solution of the sub-horizon scale linear perturbation becomes \ba \delta(Y) &=& c_{1} Y^{\fr{3 \omega_{de} -1}{6 \omega_{de}}} F [\fr{1}{2} - \fr{1}{2\omega_{de}}, \fr{1}{2} + \fr{1}{3 \omega_{de}}, \fr{3}{2} - \fr{1}{6 \omega_{de}}, -Y] \nonumber \\ && \, + \, c_{2} F[-\fr{1}{3\omega_{de}}, \fr{1}{2 \omega_{de}}, \fr{1}{2} + \fr{1}{6 \omega_{de}}, -Y] \, . \label{deltask} \ea This analytic solution does not have any physical meaning before we fix the coefficients $c_{1}$ and $c_{2}$. If we want to have the correct growing mode solution from the above analytic solution, then this solution should follow the behavior of growing mode solution at early epoch $a \simeq 0.1$. In other world, the coefficients of the $\delta$ should be fixed by using the initial conditions for the growing mode solution \be \delta(a_i) = a_{i} \hspace{0.2in} {\rm and} \hspace{0.2in} \fr{d \delta}{da} \Bigl|_{a_{i}} = 1 \, . \label{ini} \ee
\begin{center}
\begin{figure}
\vspace{1.5cm}
\centerline{
\psfig{file=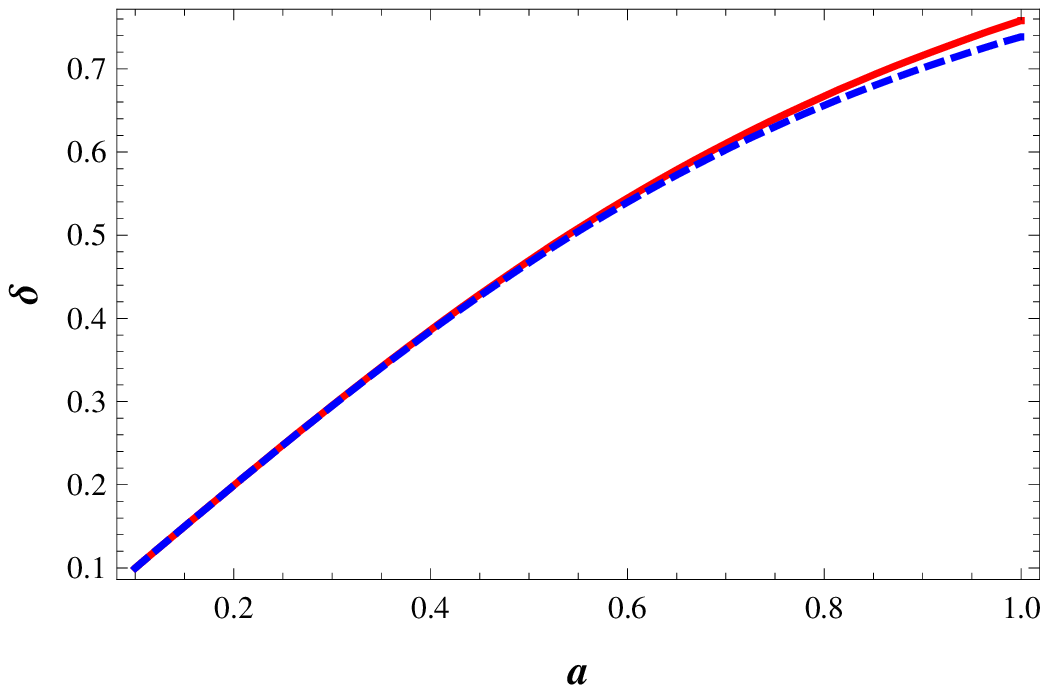, width=6cm} \psfig{file=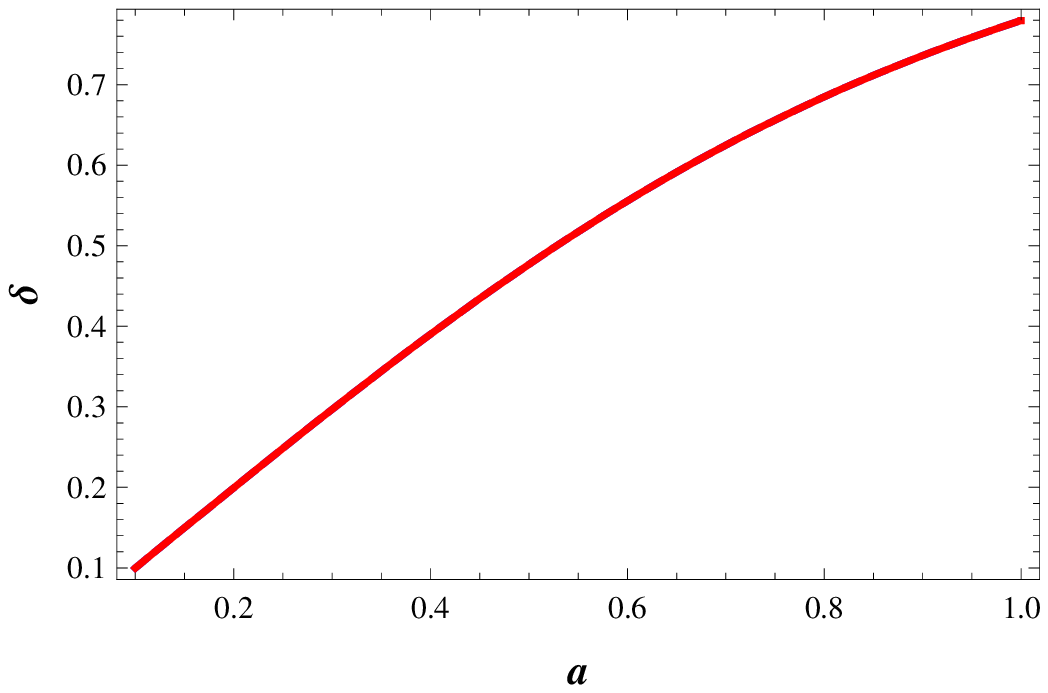, width=6cm} }
\vspace{-0.2cm}
\caption{ Evolutions of $\delta$ and $\delta^{D}$. a) When $\omega_{de} = -0.9$. b) For $\omega_{de} = -1$. } \label{fig1}
\end{figure}
\end{center}
We compare this correct analytic solution ($\delta$) with the textbook solution ($\delta^{D}$) given in the reference \cite{SK} when we use the definition of growing mode as in the text book \cite{Dodelson}. \be \delta^{D}(Y) = c_{1}^{D} Y^{-\fr{(1+\omega_{de})}{2 \omega_{de}}} \sqrt{(1 + Y)} + c_{2}^{D} Y^{\fr{(3 \omega_{de} + 1)}{3 \omega_{de}}} F[1, 1 + \fr{5}{6 \omega_{de}}, \fr{5}{2} + \fr{5}{6 \omega_{de}}, - Y ] \, . \label{deltaskD} \ee In Fig. \ref{fig1}a, we compare the behavior of them. The horizontal axis is scale factor $a$ and the vertical line indicated the amplitude of $\delta$. The solid line is the correct growing mode solution $\delta$ and the dashed line is for $\delta^{D}$. In this figure we use the initial conditions (\ref{ini}) to get the value of $c_{1}$ and $c_{2}$. For example, $(c_1, c_2)$ = $(1.088, -1.223)$ for $\delta$.  Definitely, $\delta$ is bigger than $\delta^{D}$ because $\delta^{D}$ was suppressed by the additional source term in the perturbation equation \cite{SK}. In Fig. \ref{fig1}b, we show the behaviors of $\delta$ and $\delta^{D}$ when $\omega_{de} = -1$. Definitely, two solutions are exactly matched to each other, even though the formula for two solutions look quite different for $\omega_{de} = -1$. From the initial conditions we find $(c_1, c_2) = (c_1^{D}, c_{2}^{D}) = (1.085, -0.943)$, which are used for the figure \ref{fig1}b. We can also find the behavior of the decay mode solution $\delta_{d}$ from this analytic solution after we choose the $c_{1}$ and $c_{2}$ by using the another initial conditions for decay mode solution $\delta_{d}(a_i) \propto a_{i}^{-3/2}$ and $\fr{d \delta_{d}}{d a}|_{a_i} \propto -\fr{3}{2} a_{i}^{-5/2}$. Thus, we can find the both growing and decaying mode solutions from this analytic solution $\delta$ without any ambiguity.

We need to check the growth index and growth index parameter based on this correct linear perturbation equation. Definitely, the values of these quantities are changed if we use the correct growing mode solution \cite{LN}. $\delta^{D}$ has the additional source term $- 4 \pi G (1+\omega_{de}) (1 + 3\omega_{de}) \rho_{de}$. $\delta^{D}$ oscillates at late time when $-1 < \omega_{de} < -1/3$ by including this term. Because this additional source term becomes positive and acts as a restoring force. Thus, the growth index or growth index parameters obtained from this solution are bigger than the correct values. For $\omega_{de} < -1$ or $\omega_{de} > -1/3$, this source term has the negative sign and give the additional contribution to the correct source term which should be just from the matter $-4 \pi G \rho_{m}$ if we assume that the dark energy is homogeneous. Thus, in this case $\delta^{D}$ grows faster than the correct solution $\delta$ and gives the smaller value of growth index parameter than the correct one.
\begin{center}
\begin{figure}
\vspace{1.5cm}
\centerline{
\psfig{file=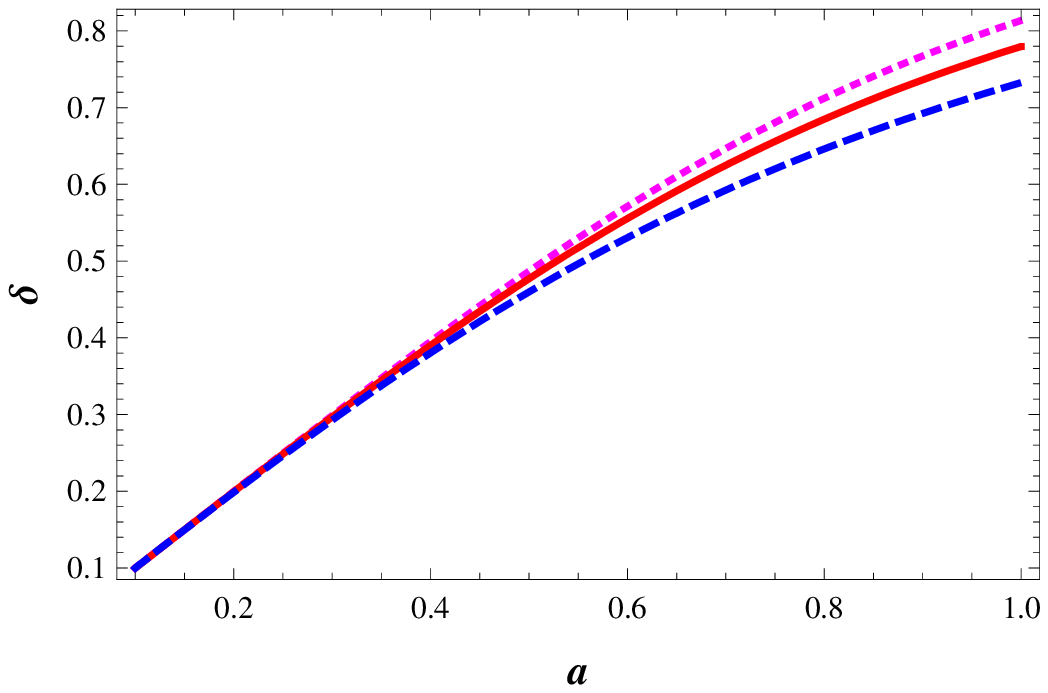, width=6cm} \psfig{file=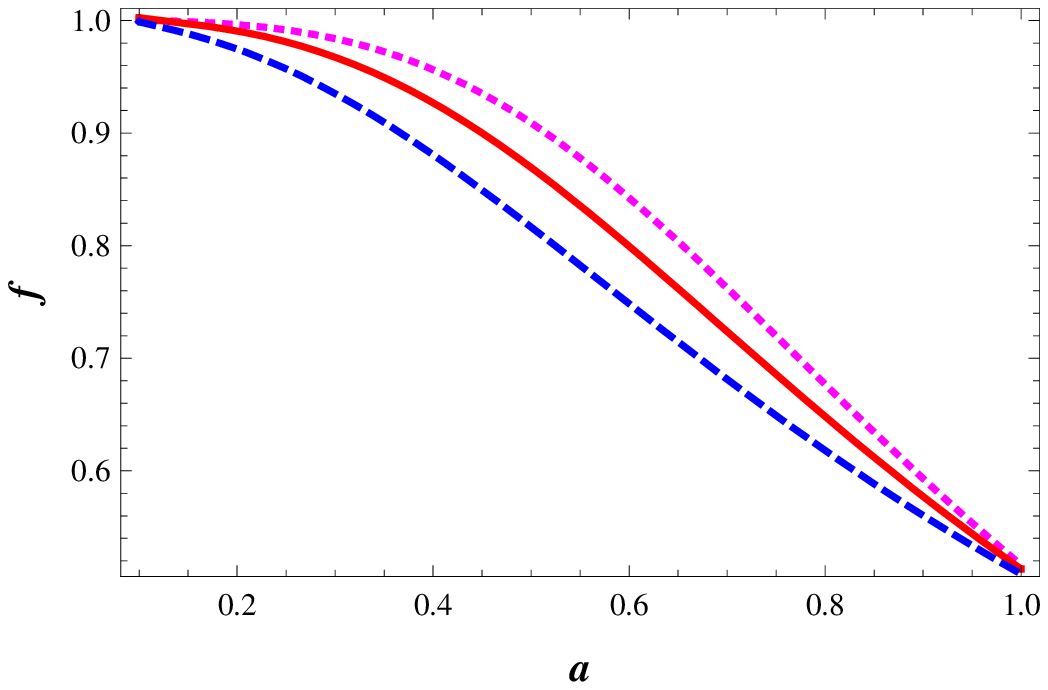, width=6cm} }
\vspace{-0.2cm}
\caption{ a) Evolution of $\delta$ for different values of $\omega_{de}$.  b) Evolutions of $f$. } \label{fig2}
\end{figure}
\end{center}
We show the behavior of $\delta$ for the different values of $\omega_{de}$ in Fig. \ref{fig2}a. Dotted, solid, and dashed lines correspond to $\omega_{de} = -1.2, -1.0$ and $-0.8$, respectively. We have more matter ratio in the past for the smaller values of $\omega_{de}$ to give the larger values of $\delta$ at present. The evolutions of the growth index $f(a) = \fr{d \ln \delta}{d \ln a}$ are depicted in Fig. \ref{fig2}b with the same notations as Fig. \ref{fig2}a.

\begin{center}
\begin{figure}
\vspace{1.5cm}
\centerline{
\psfig{file=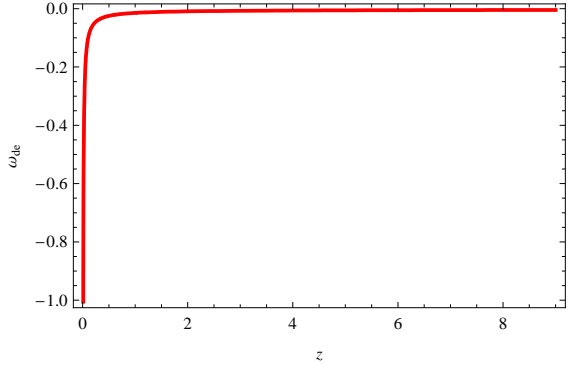, width=7.5cm}}
\vspace{-0.2cm}
\caption{ The evolution of $\omega_{de}$ in Eq. \ref{tomega}.} \label{fig3}
\end{figure}
\end{center}

We are able to extend this analytic solution in the specific form of time varying $\omega_{de}$. We choose the parametrization of $\omega_{de}$ \be \omega_{de}(a) = \omega_{1} + \fr{\omega_2}{\ln a} \label{tomega} \ee. We show the behavior of this parametrization in one specific case, $\omega_{1} = 0$ and $\omega_{2} = 0.01$ in Fig. \ref{fig3}. It rapidly changes from $z = 1$ and reaches to $-1$ at near present. Definitely, it diverges at present. One can change the slope of evolution, the present and past values from the proper values of $\omega_{1}$ and $\omega_{2}$. We want to show the possibility of extension of the exact analytic solution of the growth factor to the time varying $\omega_{de}$. We will not deep into the detail or the validity of this parametrization at this moment.
With the parametrization of Eq. (\ref{tomega}), we are able to find the exact analytic solution of $\delta$. \ba \delta(X) &=& c_{1} X^{\fr{3 \omega_{1} -1}{6 \omega_{1}}} F [\fr{1}{2} - \fr{1}{2\omega_{1}}, \fr{1}{2} + \fr{1}{3 \omega_{1}}, \fr{3}{2} - \fr{1}{6 \omega_{1}}, -X] \nonumber \\ && \, + \, c_{2} F[-\fr{1}{3\omega_{1}}, \fr{1}{2 \omega_{1}}, \fr{1}{2} + \fr{1}{6 \omega_{1}}, -X] \, . \label{deltaskt} \ea where $X = Q e^{3 \omega_{2}} a^{3\omega_{1}}$ and $Q = \fr{\Omega_{m}^{0}}{\Omega_{de}^{0}}$. It is straight forward to get this solution from the $\omega_{de}$ parametrization given n Eq. (\ref{tomega}).

The exact analytic solution provides the convenient and economic tools for probing the properties of sub-horizon scales growth factor and observational quantities related to it \cite{LN}.

We thanks Y.~Gong, E.~Linder A.~Starobinsky, S.~Varun, and A.~Wang for useful comment, K.-W.~Ng for fruitful discussion. We especially thanks S.~Habib for the correct reference and pointing out the mistake in the textbook solution.

\end{document}